# Tunable Negative Differential Resistance in Planer Graphene Superlattice Resonant Tunneling Diode

S. M. Sattari-Esfahlan[a], J. Fouladi-Oskuei and S. Shojaei

*Photonics Department, Research Institute for Applied Physics and Astronomy (RIAPA), University of Tabriz, 51665-163, Tabriz, Iran.*

Here, we study the negative differential resistance (NDR) of Dirac electrons in biased planer graphene superlattice (PGSL) and investigate the transport characteristics by adopted transfer matrix method (TMM) within Landauer-Buttiker formalism. Our model device is based on one-dimensional Kronig–Penney (KP) type electrostatic potential in monolayer graphene deposited on a substrate, where the bias voltage is applied by two electrodes in left and right. We found that, for low bias voltages, NDR is creature of breaking minibands to Wannier-Stark ladders (WSLs). At the critical bias voltage, delocalization appeared by WS states leads to tunneling peak current in current-voltage (I-V) characteristics. With increasing bias voltage, crossing of rungs from various WSL results multi-peak NDR. The results demonstrate that the structure parameters like barrier/well thickness and barrier height have remarkable effect on I-V characteristics of PGSL. In addition, Dirac gap enhances peak to valley (PVR) value due to suppressing Klein tunneling. Our results show that the tunable PVR in PGSL resonant tunneling diode (PGSLRTD) can be achievable by structure parameters engineering. NDR at ultra-low bias voltages such as 100 mV, with giant PVR of 20 is obtained. In our device, the multiple same NDR peaks with ultra-low bias voltage provide promising prospect for multi-valued memories and the low power nanoelectronic tunneling devices.

## I. INTRODUCTION

Tunneling devices are candidate for new functional devices applicable to novel integrated circuit (IC) technology in so-called "Beyond CMOS" region. Due to negative differential resistance (NDR) properties, tunnel devices show unique characteristics such as lower operating power and high speed, which are the two very substantial requirements of novel microwave amplifier [1], oscillators [2, 3], multi-valued memory [4, 5] and so on [6]. During the last decade, graphene has attracted a great deal of interest due to its superior electrical and optical properties [7-9] which becomes one of the most promising materials for nanoelectronic and nanophotonic devices [10-12]. However, graphene based structures are good candidates for tunneling devices compared to other conventional semiconductors, owing to adjustable electrical properties such as tunable energy gap and greater carrier density. Interestingly, these novel structures due to possessing higher current density and remarkable interband tunneling are useful structures for NDR based devices. Therefore, NDR features in numerous graphene-based structures have been demonstrated theoretically [13-19] and observed in experiments [20, 21]. Theoretically it has been shown that an electrostatic superlattice potential changes the electrical and transport properties of graphene [22]. Graphene superlattices (GSLs) (graphene under external periodic potential) have amazing properties; electron beam supercollimation [23] (which has not been observed in the pristine graphene), anisotropic renormalization of the carrier group velocity [24], extra Dirac points [25] and minibands [26]. In GSLs the interference of the electron wave packet, results in wannier-stark (WS) localized stetes of electron [27, 28] which separated by the Bloch oscillation frequency [29]. With applying sufficiently large bias voltage to the GSL the electronic energy band is tilted which breaks the miniband to WSL and leading to suppressing transmission. Such concept could be beneficial to designing novel NDR based devices. Moreover, the recognition of transport characterstics in biased GSL is necessary for the development of novel graphene based electronic and photonic devices.

In this work, we investigate the NDR features of an *N*- periode electrostatic monolayer PGSL with substrate induced uniform Dirac gap and calculate electronic dispersion of PGSL with Kronig-Penny model in details. We consider a linear bias voltage between left and right electrodes and calculate current-voltage (I-V) characteristics within the Landauer-Buttiker formalism. We find that the multi-peak NDR govrened by continual swithing between miniband regime to WSLs regime for high N devices where WSLs are a set of equidistant states which are confined in space [30]. Also, working window of device and additionally value of PVR could be engineered by barrier/well thicknesses, barrier height and Dirac energy gap.

## II. MODEL AND METHOD

A schematic sketch of our device is shown in Fig. 1. Metal electrodes in left and right create electrostatic linear potential across the graphene sheet by applying $V_b = V_R - V_L$ as a bias voltage. W is the PGSL width in the transversal direction and assumed large compared to the PGSL size in the propagation direction, L. $d=d_w+d_B$ is the period of PGSL where indices W/B refer to well/barrier. A graphene sheet sitting on substrate, undergoes energy gap denoted by Δ. In the other words, we consider a one-dimensional step-barrier structure in monolayer gapped graphene. The electrostatic potential $V(x)$ is shown in Fig. 1, which is assumed to be uniform along the y-direction and to vary along the x-direction, resulting in the conservation of the transverse wave vector $k_y$. The p and n region can be realized by adsorbing adatoms on graphene surface. We use Kronig-Penney model for investigation of PGSL performance. In the vicinity of the Dirac points, the GSL electronic structure can be described by the Dirac-like equation. The effective 2D Dirac Hamiltonian for a gapped GSL is written as:

$$H = -i\hbar(v_F(\sigma.p)) + eV(x)\hat{1} + \Delta\sigma_z , \qquad (1)$$

Where σ is a vector given by the Pauli matrices $\sigma = (\sigma_x, \sigma_y)$, $p = (p_x, p_y)$ is the in-plane momentum operator, $\Delta(x)$ is the graphene energy gap, $V(x)$ is position dependent electrostatic potential contains two parts: first, KP potential, $V_{KP}$ that takes as zero in well region and 200meV in barrier region. Second part indicates the external linear potential bias, $V_b$ that is taken as, $eEx$. One can write:

$$V(x) = \begin{cases} V_B(x) = V_{KP}^B - eV_{RL}x_m/L & \text{barrier} \\ V_W(x) = V_{KP}^W - eV_{RL}x_m/L & \text{well,} \end{cases} \qquad (2)$$

$x_m$ is the position of mth interface. The solution of the Dirac equation for each layer of GSL m (m = L and R for the left and right electrodes, and an m=1, 2, 3,…, n, n+1 for a total of n+1 intermediate layers) is given by the plane-wave spinors, $\psi_m(x,y) = \varphi_m(x)e^{ik_y y}$. The solution to Eq. 1 can be written as [31]:

$$\psi_1(x,y) = (A_m e^{ik_{mx}x} + B_m e^{-ik_{mx}x})e^{ik_{my}y}, \tag{3}$$

$$\psi_2(x,y) = (A_m e^{ik_{mx}x+2i\alpha} + B_m e^{-ik_{mx}x-2i\alpha})e^{ik_{my}y},$$

Where, $\psi_1(x,y)$ and $\psi_2(x,y)$ represent the components of the Dirac spinor in the $m$th monolayer graphene strip, where $A_m$ and $B_m$ are the transmission amplitudes, $\theta$ is incident angle of electron beam that is defined as angle between growth direction of planar PGSL and direction of incidence.

Based on Bloch's theorem, the electronic dispersion for periodic structure, GSL is straightforward to be obtained and the dispersion relation is given by Eq (3):

$$\cos(k_x l) = \cos(k_B d_B)\cos(k_W d_W) + \frac{k_y^2 \hbar^2 v_F^2 - (E-V_B(x))^2(E-V_W(x))^2 + \Delta^2}{\hbar^2 v_F^2 k_B k_W}\sin(k_B d_B)\sin(k_W d_W),$$

(4)

Where $k_W = (((E+eFx)^2 - \Delta^2)/\hbar^2 v_F^2 - k_y^2)^{1/2}$, $k_B = (((E-V(x)+eFx)^2 - \Delta^2)/\hbar^2 v_F^2 - k_y^2)^{1/2}$, $d_w$ and $d_B$ are the well and barrier width, respectively.

The zero-temperature current is given by [32]

$$I = I_0 \int_{\mu_L}^{\mu_R} dE |E| \times \int_{-\pi/2}^{\pi/2} \mathcal{T}(E_F, k, v_b) \cos\theta \, d\theta \tag{5}$$

Where $I_0 = 2geW/v_F h^2$, $g = 4$ is the degeneracy of electron states in graphene. A linear voltage along the x-direction defined as $eV_b = \mu_R - \mu_L$ and $\mu_R(\mu_L)$ is the bias-dependent local Fermi energy in the drain (source) electrode. $T$ is bias dependent transmission coefficient of whole structure. A detailed study along with derivation of T can be found in our previous work [33].

## III. RESULT AND DESCUSSION

Various design parameters make it possible to engineer the band structures, and subsequently transport properties of GSL. First, we consider the transmission profile of PGSLRTD for different number of *N* which is depicted in Fig. 2(a). Number of periods, *N*, forms *N* delocalized states and results flat miniband with *N*-1 spike appeared in transmission profile as can be seen in Fig. 2(a). In the absence of bias voltage minibands are formed. However, the energy location and the width of the transmission windows are independent of the N. By means of bias voltage from Fig 2.b, we observe the current initially increase with bias voltage so called miniband transport regime. With applying sufficient bias voltage WSLs are formed, which leads to spatially confined resonant modes that appear as equidistantly peaks in the transmission profile leading to resonant coupling of WSLs and subsequently resonant tunneling ( Fig. 2(b)). In the case of N=30, we see that multi-peak NDR appears with similar PVR value of 3.4. Such behavior can be implemented in multi-valued memory designing. This trend occurs in $V_b = 0.1V$, for *N*=2, $V_b$ =0.19V for *N*=5 and $V_b$ =0.1, 0.2 and 0.28V for *N*=30. Also, due to increasing discrete modes for higher *N*, the average transmission decreases.

Fig. 3 presents bias dependent transmission profile of PGSL and shows energy position of tunneling peaks labeled with TP. For *N*=2 and *N*=5 with coupling WSLs, tunnel peaks are observed only for bias window of 0.1 and 0.2V respectively (Fig. 3(a) and (b)). When bias voltage increases, resonant coupling breaks and one cannot observe any tunneling peaks in bias windows, therefore, no NDR appears in these biases. In low bias regime for *N*=30, miniband transmission is dominant that consequently creates several tunnel peaks as shown in Fig. 3(c). In fact, increasing interfaces (increasing N) enhances Fabry-Perrot like resonances in PGSL leading to increase the resonant peaks and multi peak NDR for N=30.

In Fig. 4, we present I-V characteristics of device for different $d_W$ with certain value of thin barrier. For wider wells, correspondent current peaks values are smaller than that of thinner wells. Also, the shift of peaks in I-V curve is observed towards larger biases with decreasing well width. This behavior can be attributed by red shifting of resonant energies with increasing well width. In the case of wider wells, lower biases are needed to align the resonant states, so current reach the peak more quickly than that of thinner wells. For thinner wells, I-V curve gradient is remarkably big, therefore current peak are greater than those of wider wells. However, due to adjacent subbands inside wide wells, band to band tunneling is noticeable which increases valley current and extremely decreases PVR. In contrast to this finding, increasing barrier width doesn't change the current trend (not shown here).

Other parameter that influences I-V characteristics in PGSL barrier height that in our work is induce by adatoms create p/n regions (Fig. 1). Fig. 5 depicts the effect of barrier height on I-V curve. It is clear that current peak is decreased by decreasing barrier height. From transmission spectrum it was determined that shorter barrier shift the transmission gap to lower energies, that subsequently weakens the tunneling probabilities results in decreasing the current. Also, from Fig. 5 it can be seen that decreasing barrier height shifts the current peak towards lower biases.

Dirac gap induced by substrate is essential to eliminate Klein tunneling, which enhances gap in transmission profile and confinement of the Dirac electrons in graphene. In the other words, energy gap destroys the perfect reflection for Dirac electrons traversing the barrier with $k_y = 0$. These two latter cases are determinant factors to NDR. Putting graphene on substrate is a simple and powerful way to generate Dirac gap [34]. Hence, in Fig. 6 we show Dirac gap role in NDR mechanism. With modifying Dirac gap value by substrate we found that PVR value of 1.6 at $\Delta= 0.05\ eV$ increases exponentially to value of 20 at $\Delta= 0.15\ eV$ whereas, the position of peak and valley current are nearly fixed in 0.2 V and 0.25 V, respectively. Furthermore, we found that from transmission spectrum this increment in PVR is affected by increasing transmission gap.

**IV. CONCLUSION**

In summary, we have investigated the influence of structural parameters on transport characteristics of PGSL. It is shown that the NDR behavior strongly dependent on device parameters. Values of 20 obtained for PVR in very low bias regime for *Δ=100 meV* and $d_W = 20, d_B = 10\ nm$. Our proposed device along with theoretical study pave a novel way in the control of graphene based NDR devices such as resonant tunneling diode and electronic filters.

**Figure caption:**

Fig. 1: (Color online) (a) Band structure of biased PGSL. Electron beam incident angle is also shown. (b) Schematic view of simulated device where in the graphene sheets are deposited on a substrate. Grphene sheet is composed of *p* and *n* doped region as periodic barrier and wells and $V_b$ is bias voltage between left and right electrode.

Fig. 2: (Color online) a. Transmission spectrum of Dirac electrons through PGSL for *N*=2, *N*=5 and *N*=30 periods. Vb=0, $d_W = 20, d_B = 10$ nm.
b. I-V characteristics for different period number of PGSL with $d_W = 20$ and $d_B = 10$ nm.

Fig. 3: (Color online) Transmission spectrum of PGSL at the various bias voltages (a) N=2, (b) N=5, (c) N=30. Black dashed curves are transmission coefficient at Vb= 0. TP label shows the tunneling peaks. Vertical lines indicate width of bias windows.

Fig. 4: (Color online) I-V characteristics of PGSL for different values of $d_W$ . Δ=100 meV, UB=...meV  $d_B = 10$ nm, N=5.

Fig. 5: (Color online) I-V characteristics of PGSL for different values of $U_B$. *Δ=100 meV* , $d_W = 20, d_B = 10\ nm$.

*N*=5.

Fig. 6: (Color online) I-V characteristics of PGSL for different values of *UB=...meV* and $d_W = 20, d_B = 10\ nm$. *N*=5.

Fig. 1:

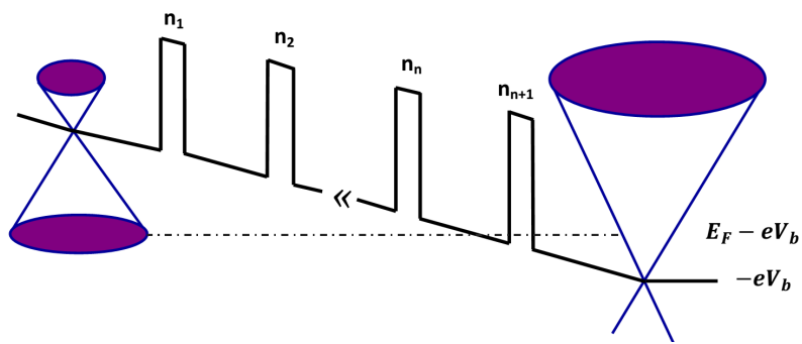

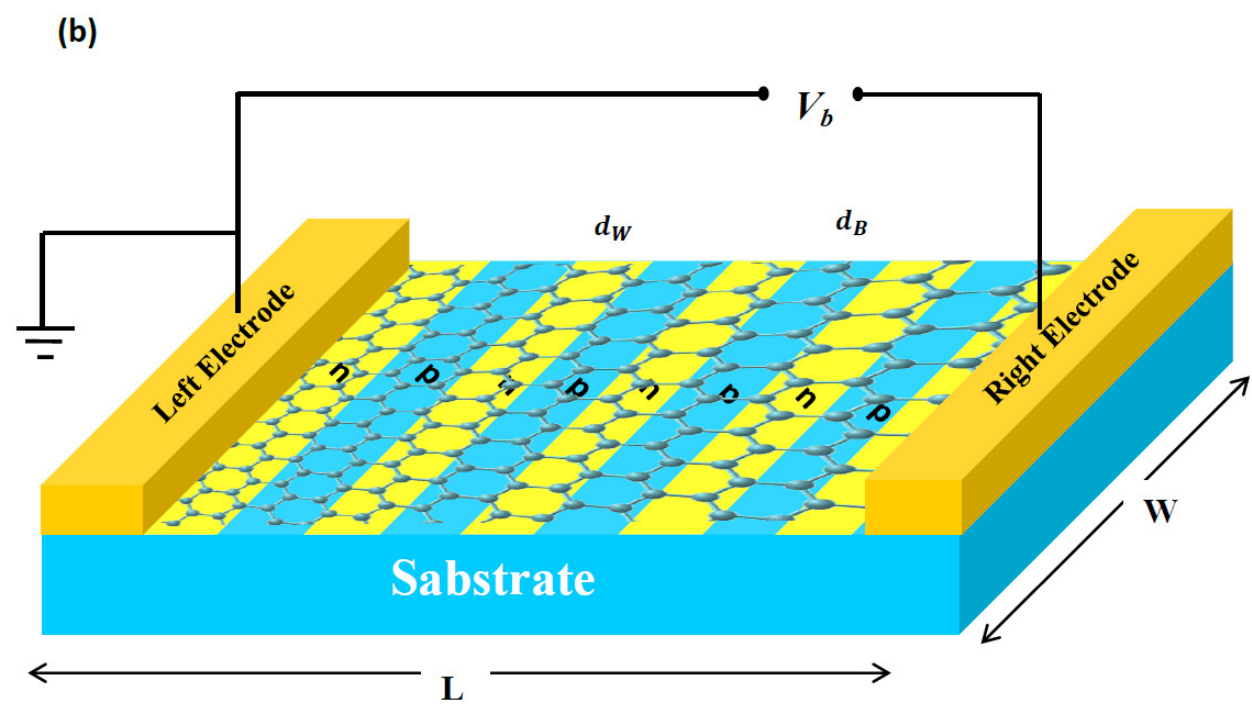

Fig. 2:

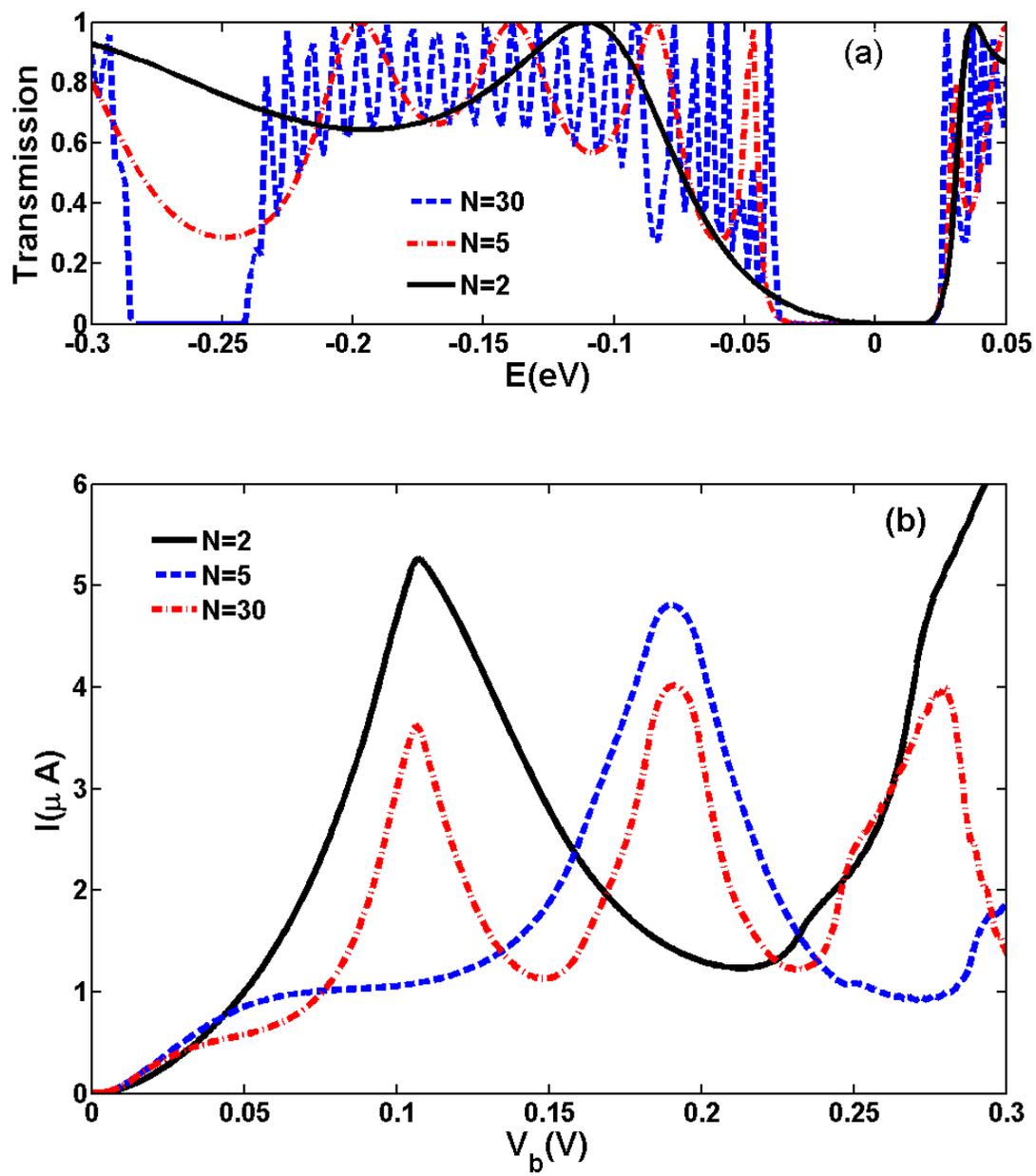

Fig. 3:

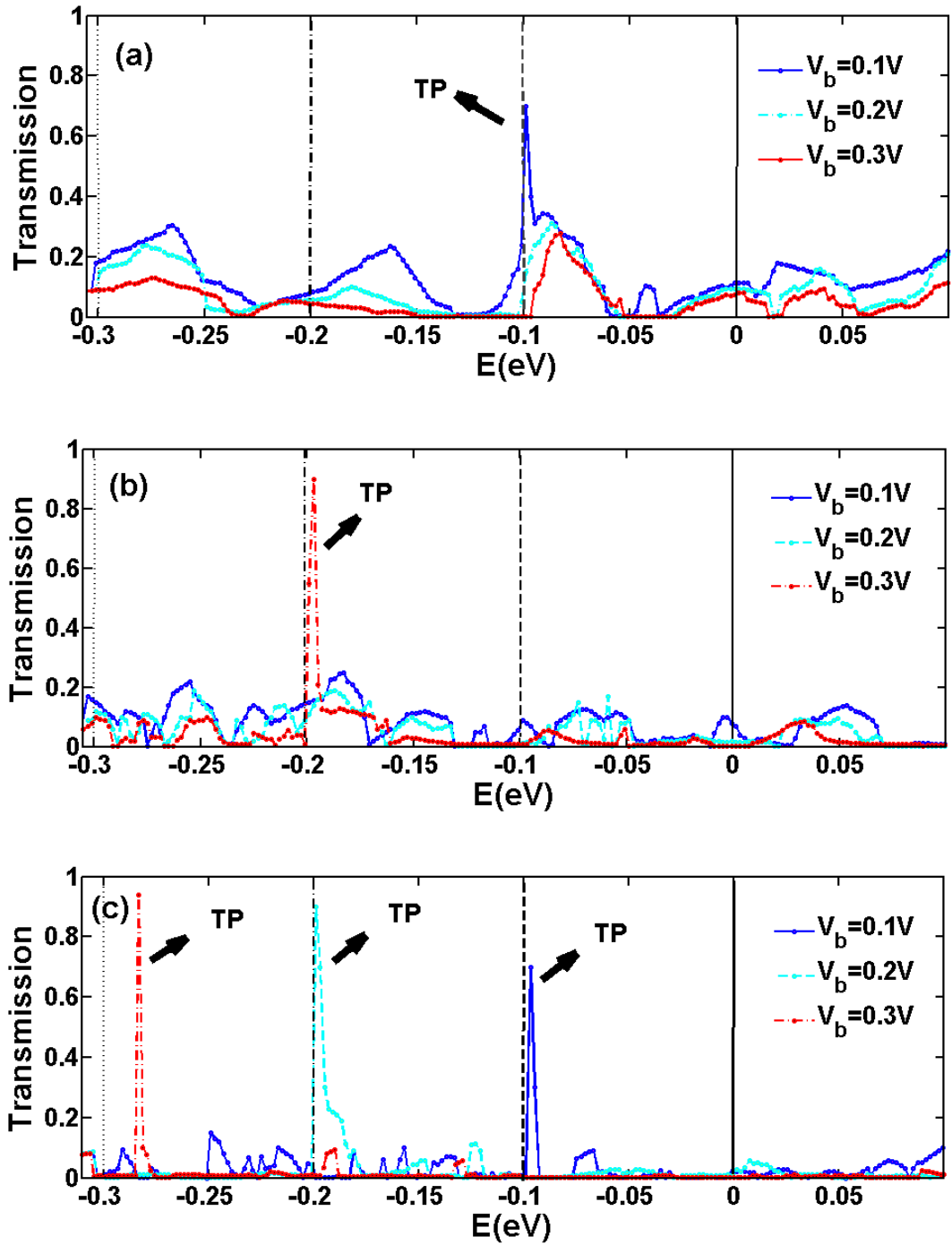

Fig. 4:

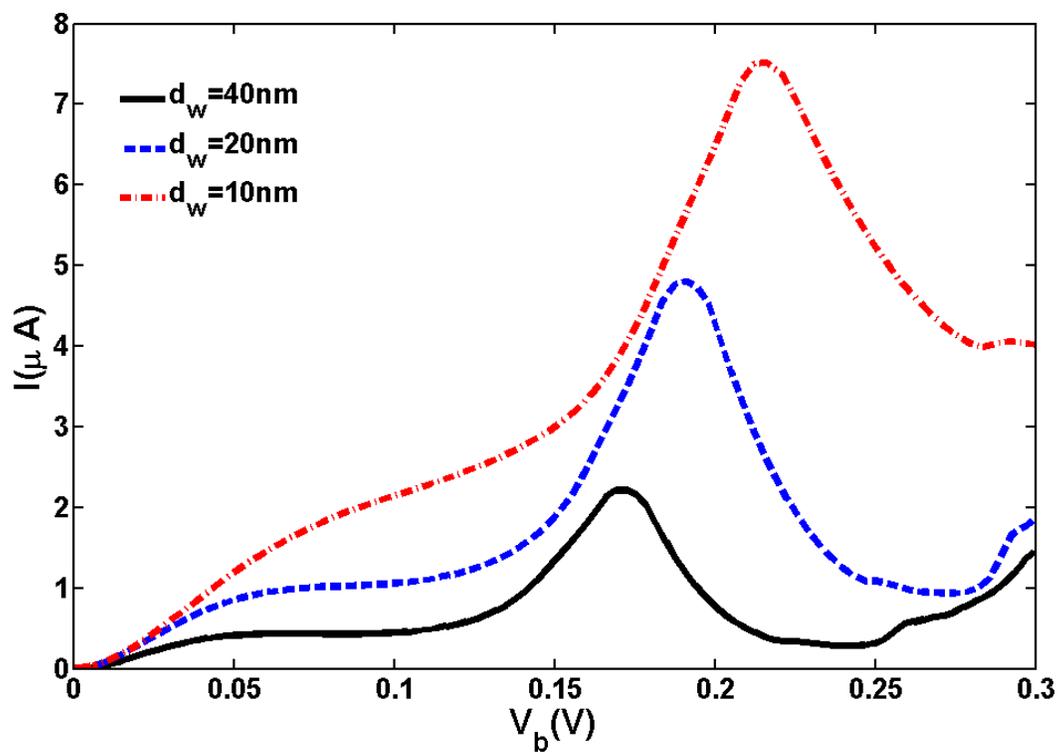

Fig. 5:

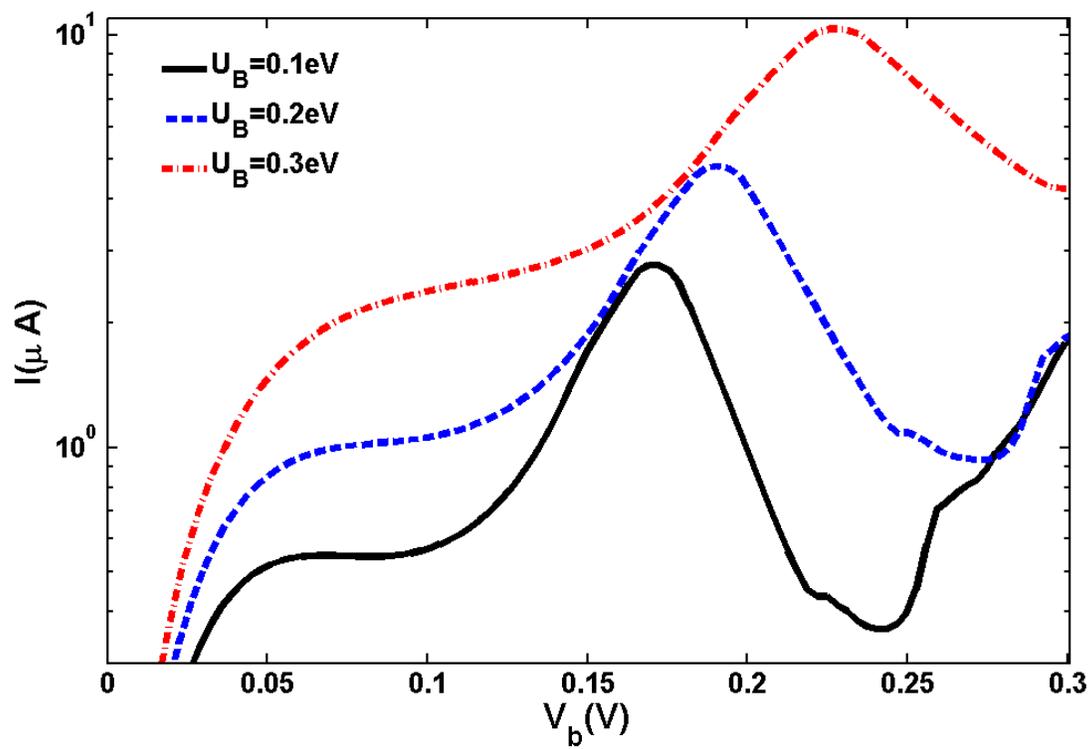

Fig. 6:

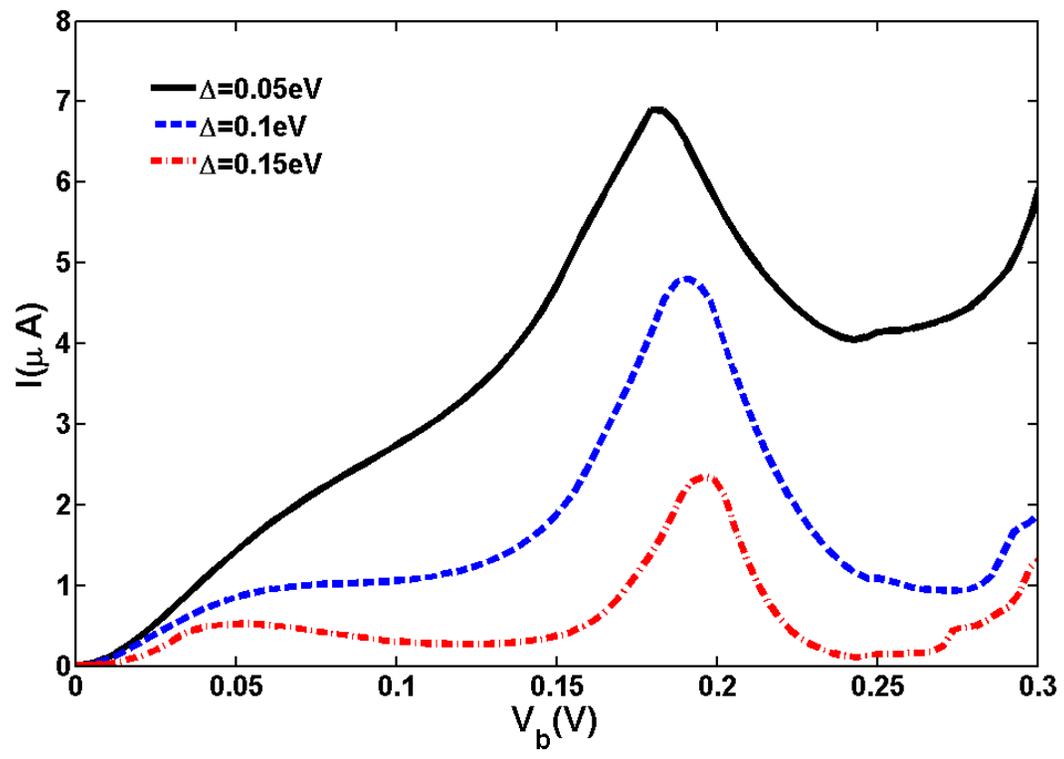